\documentclass[showpacs,aps,pra,reprint,amsmath,amssymb]{revtex4-1}
\usepackage{amsmath}
\usepackage{epsfig}
\usepackage{subfigure}
\usepackage{graphicx}
\usepackage{dcolumn}
\usepackage{stmaryrd}
\usepackage{mathrsfs}
\usepackage{pifont}
\usepackage{amsthm}
\usepackage{amssymb}
\usepackage{bm}
\usepackage{latexsym}
\usepackage{hyperref}
\usepackage{color}
\usepackage{amsfonts}%
\providecommand{\U}[1]{\protect\rule{.1in}{.1in}}
\newcommand{\la}{\langle}
\newcommand{\ra}{\rangle}
\newcommand{\beq}{\begin{equation}}
\newcommand{\eeq}{\end{equation}}
\newcommand{\beqa}{\begin{eqnarray}}
\newcommand{\eeqa}{\end{eqnarray}}

\newcommand{\+}{^\dagger}

\newcommand{\om}{\omega}

\newcommand{\pa}{\partial}
\newcommand{\non}{\nonumber}
\newcommand{\dg}{\mathcal{D}_g[\bx]}
\newcommand{\rd}{{\rm d}}
\newcommand{\ri}{{\rm i}}
\newcommand{\rs}{{\rm S}}
\newcommand{\rr}{{\rm R}}
\newcommand{\re}{{\rm e}}
\newcommand{\rt}{{\rm tot}}
\newcommand{\eq}{\equiv}
\newcommand{\sq}{\sqrt}
\newcommand{\ora}{\overrightarrow}
\newcommand{\ola}{\overleftarrow}

\newcommand{\bn}{{\bm n}}
\newcommand{\bmm}{{\bm m}}
\newcommand{\bl}{{\bm l}}
\newcommand{\bx}{{\bm \xi}}
\newcommand{\mm}{{\mathcal M}}
\newcommand{\fr}{\frac}
\newcommand{\idg}{\int \mathcal D_g [\bm\xi]}

\begin{document}
\title{Non-Markovian Fermionic Stochastic Schr\"{o}dinger Equation for Open System Dynamics}
% \author{Wufu Shi\footnote{Email: wshi1@stevens.edu},  Xinyu Zhao\footnote{Email: xzhao1@stevens.edu}, and Ting Yu\footnote{Email: ting.yu@stevens.edu}%
% %EndExpansion
% }
\author{Wufu Shi}
\email{Email:wshi1@stevens.edu}
\author{Xinyu Zhao}
\email{Email:xzhao1@stevens.edu}
\author{Ting Yu}
\email{Email:ting.yu@stevens.edu} 

\affiliation{Center for Controlled Quantum Systems and the Department of Physics and Engineering Physics, Stevens Institute of
Technology, Hoboken, New Jersey 07030, USA}
\date{\today }

\begin{abstract}
This paper presents an exact Grassmann stochastic Schr\"{o}dinger
equation for the dynamics of an open fermionic quantum system coupled to a
reservoir consisting of a finite or infinite number of fermions. We use this
stochastic Schr\"{o}dinger equation as a generic open system tool to derive the exact master equation for
an electronic system strongly coupled to fermionic reservoirs. The generality and applicability 
of this Grassmann stochastic approach are justified and exemplified by several
quantum fermionic system problems concerning quantum coherence coupled 
to vacuum or finite-temperature fermionic reservoirs.  Our studies show
that the quantum coherence property of a quantum dot system can be
profoundly modified by the environmental memory.  

\end{abstract}

\pacs{03.65.Yz, 42.50.Lc, 05.40.-a}
\maketitle

\section{introduction}
%------------------------------------------------------------------------------------------------------------------------
Quantum dynamics of quantum systems coupled to
fermionic  or bosonic  environments has recently attracted
wide-spread interest in quantum open systems, quantum dissipative systems, quantum transport, 
quantum computing, and nanoscience \cite{xx,yy,nnn,mmm}.  For example, the size reduction 
of quantum devices in microelectronics requires controllable systems consisting of 
only a few electrons, where quantum coherence and quantum interference become dominant.
In addition, quantum dots coupled to electrons of a metal is an interesting
setup in quantum information processing where the quantum coherence of
qubits is of essential importance \cite{cc}. The non-Markovian open systems
arise in many important situations such as the strongly coupled
system-environment, structured environment,  time-delayed external control etc. \cite{Hu0}.
Intuitively speaking, while the Markov evolution is an irreversible process, in the case of non-Markovian dynamics, 
the system energy (or phase information) dissipated into the environment may come back to the system in a finite time \cite{Breuer2009}.
For open systems  immersed in a bosonic environment, apart from the master equation and path
integral approaches \cite{Feynman-Vernon,Leggett,Hu1},  a versatile stochastic formalism for 
quantum open system dynamics was developed to provide a powerful tool in studying
quantum systems in a non-Markovian regime
\cite{Diosi1,Diosi2,Strunz,YDGS99,Jing,Strunz-Yu2004,Yu2004}. 
Such a stochastic pure state approach has several advantages in numerical simulations, perturbation
and the derivation of the corresponding master equations.
For a Markov environment (bosons or fermions), both the quantum state diffusion equations
\cite{Gisin,Plenio} and Lindblad master equations \cite{Quantum Noise} can be used to
describe quantum dynamics of the system of interest.  Several important theoretical tools
in dealing with nonequilibrium fermionic systems have been developed including nonequilibrium 
Green's function (NEGF) theory, fermionic path integral etc \cite{xx,Meir,Meir2,Jauho,Wingreen,Fransson,Zwanzig}. However,  for a generic
non-Markovian fermionic environment where the system-environment coupling
is not weak or the environment cannot be treated as a broadband reservoir \cite{mmm,Meir,Meir2,Jauho,Wingreen,Fransson,Zwanzig,Ting,ZhangDQD,Wiseman}, 
establishing a stochastic theory analogous to the non-Markovian quantum state diffusion equation \cite{Strunz} 
is a long standing problem. 

The purpose of this paper is to develop a general 
non-Markovian stochastic theory of electronic systems coupled to a fermionic environment.
The theory developed here is versatile enough to deal with a  wide spectrum of open system problems  ranging from a
``small"  environment (one fermion or a few fermions) to a large environment consisting of an infinite number of fermions
irrespective of the details of spectral distribution of the fermionic environment.
As an illustration of the power of the stochastic approach developed here, we derive several exact
master equations governing the reduced density operators of the electronic
systems coupled to vacuum and finite-temperature reservoirs.

The paper is organized as follows. In Sec.~\ref{Model}, we establish the fermionic stochastic Schr\"{o}dinger equation (SSE) for a class of open system models
consisting of an electronic system coupled to a fermionic reservoir. In particular, we show how to derive the time-local SSE. We introduce a new type of Novikov
theorem emerged from the Grassmann noise. We demonstrate the derivation of the corresponding exact master equation from the SSE. 
In Sec.~\ref{example1}, a many-fermion model is considered. We show explicitly that the exact fermionic SSE and the corresponding master equation can be
established.   In Sec.~\ref{example2},  we establish the finite temperature fermionic SSE through a Bogoliubov transformation, and we provide an 
explicit construction of the exact time-local master equation as well as the so-called $\hat{Q}$ operator for this model. Then, in Sec.~\ref{example3}, the finite temperature model is generalized into a more realistic case consisting of  double quantum dots coupled to two fermionic reservoirs (source and drain). 
Both the time-local fermionic SSE and the exact master equation are derived. The numerical
 simulations of the double quantum dots based on the exact master equation are provided.  Finally, 
 we conclude the paper in Sec.\ref{conclusion}.  Some details about the Grassmann noise, the fermionic SSE, the $\hat{Q}$ operator,  a proof of the Novikov theorems and the Heisenberg operator approach are left to Appendix.

\section{Fermionic stochastic Schr\"{o}dinger equation and non-Markovian master equation}
\label{Model}
\subsection{Model}
To begin with, we consider a simplified model involving an electronic system in contact
with a single fermionic reservoir, where the system anti-commutes with the bath
\cite{XinyuFB}. The generalization to a more physically interesting model with 
two reservoirs (\emph{e.g.}, source and drain) can be established in a similar way. 
With necessary modifications, the  formalism is versatile enough to deal with stochastic gate 
potentials and nonlinear couplings. The total Hamiltonian for the system plus environment may
be written as \cite{xx},
\begin{equation}
\hat{H}_{\mathrm{tot}}=\hat{H}_{\mathrm{S}}+\hat{H}_{\mathrm{R}}+\hat
{H}_{\mathrm{I}}, \label{Htot}%
\end{equation}
where $\hat{H}_{\mathrm{S}}$ is the Hamiltonian of the electronic system in
the absence of the environment, $\hat{H}_{\mathrm{R}}$ is the Hamiltonian for a
fermionic reservoir; $\hat{H}_{\mathrm{R}}=\sum_{\mathbf{k\alpha}}\hbar
\omega_{\mathbf{k}}\hat{b}_{\mathbf{k\alpha}}^{\dag}\hat{b}_{\mathbf{k\alpha}%
}$ where $\hat{b}_{\mathbf{k\alpha}}^{\dag},\hat{b}_{\mathbf{k\alpha}}$ are
the fermionic creation and annihilation operators $\{\hat{b}_{\mathbf{k\alpha
}},\hat{b}_{\mathbf{k^{\prime}\alpha^{\prime}}}^{\dag}\}=\delta
_{\mathbf{k\alpha,k^{\prime}\alpha^{\prime}}}$, and the interaction
Hamiltonian $\hat{H}_{\mathrm{I}}$ is given by
\begin{equation}
\hat{H}_{\mathrm{I}}=\hbar\sum_{\mathbf{k\alpha}}(t_{\mathbf{k\alpha}}\hat{L}^\dagger \hat{b}%
_{\mathbf{k\alpha}} + t_{\mathbf{k\alpha}}\hat{b}^\dagger_{\mathbf{k\alpha}}%
\hat{L}), \label{Hint}%
\end{equation}
where $\hat{L}$ is the system coupling operator and $t_{\mathbf{k\alpha}}$ are
the coupling constants. Note that $\hat{H}_{\mathrm{S}}$ is an arbitrary
Hamiltonian operator that may contain interaction terms for the system
particles (\emph{e.g.}, Coulomb interactions between two electrons). The
coupling operator $\hat{L}$ may, in general, be represented by a set of
fermionic operators which are coupled to all the participating external agents
such as the source and drain reservoirs.

The purpose of this paper is to develop a systematic stochastic theory for
the models described by Eq. (\ref{Htot}) and (\ref{Hint}), those are relevant to quantum open systems,
non-equilibrium statistical mechanics, path-integral theory and quantum
devices based on quantum dots and mesoscopic electronics
\cite{Feynman-Vernon,Leggett,Hu1,Hu0,Wiseman}.
In the framework of the stochastic Schr\"{o}dinger equation (SSE) for a bosonic bath, the state of the open
quantum system is described by a stochastic pure state, which is generated by
a complex Gaussian stochastic process. For the fermionic environment considered in this paper, 
similar to the fermionic
path integral, the fermionic stochastic theory will involve a Grassmann
Gaussian stochastic process.  Remarkably, we show that the reduced density matrix of the system of
interest can be reconstructed from the pure states by taking the statistical
mean over the Grassmann noise \cite{XinyuFB}. As such, in principle,  the exact master
equation governing the reduced density operator for the open system can be recovered 
from the SSE, as illustrated by several physically interesting models below.

\subsection{Fermionic stochastic Schr\"{o}dinger equation}
%----------------------------------------------------------------------------------------------------
In this subsection, we will establish the fermionic stochastic Schr\"{o}dinger equation (SSE) for an open electronic
system coupled to a fermionic reservoir.  Consider the model described by the total Hamiltonian in
Eq.~(\ref{Htot}), in the interaction picture with respect to the fermionic
reservoir,  it becomes (setting $\hbar=1$), 
\begin {eqnarray}
  \hat{H}_{\mathrm{tot}}^{I}(t)=\hat{H}_{\mathrm{S}}+(\sum_{j}t_{j}\hat{L}^{\dagger}\hat{b}_{j}e^{-\ri\omega_{j}t}+\mathrm{h.c.}),
\end {eqnarray}
here the subscript $\mathbf{k\alpha}$
is suppressed as $j$. In order to trace out the environmental variables, we
introduce the fermionic coherent states $|\bm{\xi}\rangle$ which is defined as
\begin {eqnarray}
|\bm{ \xi}\rangle \equiv \prod_k (1-\xi_{k} \hat{b}_k^{\dagger})|{\rm vac}\rangle_{\rm R}. 
\end {eqnarray}
And this state satisfies  $\hat{b}_{j}|\bm {\xi}\rangle = \xi_{j}|\bm {\xi}\rangle$. Here, $\xi_{j}$ is a
Grassmann variable, satisfying $\{\xi_i,\xi_j\}=\{\xi_i^*,\xi_j\}=0$, and $\{\xi_i,\hat
{b}_j\}=\{\xi_i,\hat{b}_j^{\dag}\}=0$ \cite{Berezin,Glauber1999}. As shown below,
the derivations and results for fermionic SSE are more complex than the
bosonic case. Using the fermionic coherent state, we define 
\begin {eqnarray}
 |\psi_{t}(\bm{\xi}^{\ast})\rangle \equiv \langle\bm{\xi}|\re^{\ri\hat{H}_{\mathrm{R}}t} \re^{-\ri\hat{H}_{\mathrm{tot}}t}|\psi_{\mathrm{tot}}(0)\rangle, \label{psit}
\end {eqnarray}
where $|\psi_\rt(0)\ra$ is the total initial state of both
system and bath, and we assume the bath is initially prepared in vacuum
state, \emph{i.e.} $|\psi_\rt(0)\ra=|\psi_\rs(0)\ra\wedge|\mathrm{vac}\ra_\rr$ (``$\wedge$''  stands for an antisymmetrized wave function). The thermal state case will be discussed 
later. Taking the time derivative on the both sides of Eq.~(\ref{psit}),  one obtains%
\begin{align}
& \pa_t|\psi_{t}(\bm{\xi}^*)\ra \non \\
=& -\ri\la{\bm\xi}|\hat{H}_\rt^{I}(t)|\psi_\rt^I(t)\ra \non \\
=& -\ri [\hat{H}_\rs + \sum_k (t_k\hat{L}\+\ora{\pa}_{\xi_{k}^*}\re^{-\ri\om_{k}t} + t_k\xi_k^*\hat{L}\re^{\ri\om_{k}t})]|\psi_{t}(\bm{\xi}^*)\ra \non \\
=& (-\ri\hat{H}_\rs-\hat{L}\+\int_0^t \rd s\,\sum_k\frac{\pa\xi_t}{\pa\xi_k}\frac{\pa\xi_s^*}{\pa\xi_k^*}\ora{\delta}_{\xi_s^*}-\hat{L}\xi_t^*)|\psi_t(\bm{\xi}^*)\ra \non \\
=& [-\ri\hat{H}_\rs -\hat{L}\xi_t^* -\hat{L}\+ \int_0^t \rd s\, K(t,s) \ora{\delta}_{\xi_{s}^*}]|\psi_t(\bm{\xi}^*)\ra,\label{SSE}%
\end{align}
where $\xi_t^{\ast} \equiv-\ri\sum_{k}t_{k} \re^{\ri\omega_{k}t}\xi_k^{\ast}$
is the Grassmann Gaussian noise, $\overrightarrow{\delta}_{\xi_s^{\ast}}$ is the left functional derivative
with respect to $\xi_s^*$, and the explicit form of function $K(t,s)$ is 
\begin {eqnarray}
 K(t,s) \equiv \sum_{k}\frac{\partial \xi_t}{\partial \xi_k}\frac{\partial
\xi_s^{\ast}}{\partial \xi_k^{\ast}}=\sum_{k}|t_{k}|^{2} \re^{-\ri\omega_{k}(t-s)}.
\end {eqnarray}

The Grassmann Gaussian process is defined by
\begin {eqnarray}
 & & \mathcal{M} [\xi_k ]\equiv \int \prod_k \rd\xi_{k}^{\ast}\cdot \rd\xi_{k}\, \re^{-\xi_{k}^{\ast}\cdot \xi_{k}} \, \xi_k =0, \non \\
 & & \mathcal{M} [\xi_k \xi_k^*]\equiv \int \prod_k \rd\xi_{k}^{\ast} \cdot \rd\xi_{k}\, \re^{-\xi_{k}^{\ast}\cdot \xi_{k}}\, \xi_k \xi_k^*=1,
\end {eqnarray}
 where ``$\mathcal{M}$'' stands for the statistical mean over the random Grassmann
variables ``$\xi_k$''. It is easy to check that the mean and the correlation function are given by: $\mathcal{M} [\xi_t ]= \mathcal{M} [\xi_t^*]=0$ 
and $ \mathcal{M} [\xi_t \xi_s^*]=K(t,s)$, respectively.
Note that our fermionic SSE (\ref{SSE}) is applicable to an arbitrary correlation function including both  Markov and 
non-Markovian environments. Unlike the complex Gaussian noise used in bosonic case, the Grassmann Gaussian
noise is a non-commutative noise at different times reflecting a fundamentally distinctive feature arising from the 
fermionic environment. Our fermionic stochastic Schr\"{o}dinger equation is expected to have a close connection 
with the fermionic path integral as shown in the case of bosonic case \cite{Strunz1996,Wufu}.

The fermionic SSE  (\ref{SSE}) can be written in a more compact form, 
\begin{equation}
\partial_{t}|\psi_{t}(\bm{\xi}^*)\rangle=-\ri\hat{H}_{\mathrm{eff}}%
|\psi_{t}(\bm{\xi}^{\ast})\rangle, \label{SSE1}%
\end{equation}
where the effective Hamiltonian is given by,%
\begin{equation}
\hat{H}_{\mathrm{\mathrm{eff}}}=\hat{H}_{\mathrm{S}}-\ri\hat{L}\xi_t%
^{\ast}-\ri\hat{L}^{\dagger}\int_{0}^{t}\rd s\, K(t,s)\overrightarrow{\delta
}_{\xi_s^{\ast}}, \label{SSE2}%
\end{equation}

Eq.~(\ref{SSE}) or (\ref{SSE1}) may serve as a fundamental equation for open fermionic
systems coupled to a fermionic environment.  Our stochastic method will provide 
a new insight into the individual physical process described by the SSE. 
Although the stochastic method is fundamentally equivalent to density matrix or NEGF formalism,
it can be advantageous over the density operator and NEGF in several interesting cases such as fast tracking of information 
for quantum coherence  and entanglement \cite{Corn}.  Moreover, it is known that perturbative master equations typically
lead to unphysical effect such as violation of positivity, however,  the stochastic equation can yield a 
systematic perturbative method that can be implemented numerically \cite{YDGS99}.
As an illustration of the power of the stochastic approach developed here, we derive several exact
master equations governing the reduced density operators of the electronic
systems coupled to vacuum and finite-temperature reservoirs.

Crucial to the practical  applications of Eq.~(\ref{SSE1}) is to express the Grassmann functional
derivative under the memory integral in Eq. (\ref{SSE2}) in terms of system
operators \cite{Diosi2,Strunz,YDGS99,Jing,Strunz-Yu2004,Yu2004}. In order to
calculate the functional derivative in the stochastic Schr\"{o}dinger
equation, we introduce an operator called the fermionic $\hat{Q}$ operator
(similar to the $\hat{O}$ operator in bosonic case) as
\begin{equation}
\hat{Q}(t,s,\bm{\xi}^{\ast})|\psi_{t}(\bm{\xi}^{\ast})\rangle \equiv \overrightarrow
{\delta}_{\xi_{s}^{\ast}}|\psi_{t}(\bm{\xi}^{\ast})\rangle.
\end{equation}
With this definition, the effective
Hamiltonian in Eq.~(\ref{SSE2}) can be written as,%
\begin{equation}
\hat{H}_{\mathrm{eff}}=\hat{H}_{\mathrm{S}}-\ri\hat{L} \xi_%
t^{\ast}-\ri\hat{L}^{\dagger}\bar{Q}, \label{SSE3}%
\end{equation}
where 
\beq
\bar{Q}(t,\bm{\xi}^{\ast}) \equiv \int_{0}^{t}\rd s\, K(t,s)\hat{Q}%
(t,s,\bm{\xi}^{\ast}).
\eeq 

The fermionic stochastic Schr\"{o}dinger equation (\ref{SSE1}) is derived directly from the
microscopic Hamiltonian without any approximation. It should be emphasized
that the system Hamiltonian $\hat{H}_{\mathrm{S}}$ and the coupling operator
$\hat{L}$ are entirely general. The evolution of the electronic system is
governed by the anti-commutative stochastic differential equation (\ref{SSE1}). Although the
mathematical form of the equation (\ref{SSE1}) is similar to the non-Markovian quantum
state diffusion equation in the bosonic case, the behavior
of the fermionic SSE can be very different due to the fermionic features of the environment
\cite{XinyuFB}.  Moreover, the Grassmann stochastic process has brought about
several new features in dealing with the fermionic SSE such as a new type of
Novikov theorem (see Appendix  D).

From the consistency condition for the fermionic SSE 
\begin {eqnarray}
 \overrightarrow{\delta}_{\xi_{s}^{\ast}}\partial_{t}|\psi_{t}(\bm{\xi}^{\ast})\rangle
=\partial_{t}\overrightarrow{\delta}_{\xi_{s}^{\ast}}|\psi_{t}(\bm{\xi}^{\ast})\rangle,
\end {eqnarray}
 the fermionic $\hat{Q}$ operator can be shown to satisfy the
following equation (see Appendix B),
\begin{equation}
\partial_{t}\hat{Q}=-\ri[\hat{H}_{\mathrm{\mathrm{eff}}},\hat{Q}%
]-\ri\overrightarrow{\delta}_{\xi_{s}^{\ast}}(\hat{H}_{\mathrm{eff}}%
-\hat{H}_{\mathrm{S}}). \label{EQO}%
\end{equation}
Once the fermionic $\hat{Q}$ operator is determined, the SSE can be cast into
a time-local stochastic equation with the Grassmann type noise. 

\subsection{Non-Markovian master equation}
Note that the
reduced density operator for the open fermionic system can be obtained by
taking the statistical average over all the Grassmann quantum trajectories 
which are the solutions to the SSE (\ref{SSE1}),
\begin{align}
\hat{\rho}_{r}  &  =\int\prod_k \rd\xi_{k}^{\ast}\cdot \rd\xi_{k}%
\re^{-\xi_{k}^{\ast}\cdot\xi_{k}} \hat{P},\\
\hat{P}  &  =|\psi_{t}(\bm{\xi}^{\ast})\rangle\langle\psi
_{t}(-\bm{\xi})|, \label{P}%
\end{align}
and in the rest of the paper, we will use the shorthand notations $\mathcal{D}_{g}[\bm{\xi}] \equiv \prod_k \rd\xi_{k}^{\ast}\cdot \rd\xi_{k}%
\re^{-\xi_{k}^{\ast}\cdot\xi_{k}}$ and  $|\psi_{t}\rangle\equiv|\psi_{t}(\bm{\xi}^{\ast})\rangle$, $|\psi_{t}^-\rangle\equiv|\psi_{t}(\bm{-\xi}^{\ast})\rangle$ 
to represent the Grassmann Gaussian measure and the quantum trajectories, respectively  (for more details, see Appendix C). Then taking the time
derivative on
\begin {eqnarray}
 \hat{\rho}_{r}  &  =\int\mathcal{D}_{g}[\bm{\xi}]|\psi_{t}\ra \la\psi_{t}^-|,
\end {eqnarray}
and substituting the fermionic SSE (\ref{SSE1}) into it,  we can get the equation of motion for the reduced density operator, 
 \begin {eqnarray}
      \pa_t \hat{\rho}_r &=& -\ri[\hat{H}_\rs , \hat{\rho}_r]  + \int \dg \{ (-\hat{L}\+ \bar{Q} - \hat{L}\xi^*_t )\hat{P} \non \\
      & & +\hat{P}(-\bar{Q}_- \+ \hat{L} + \xi_t\hat{L}\+ ) \}, \label{rho3}
 \end {eqnarray}
where $\bar{Q}_-$ is a short hand notation of $\bar{Q}(-\bx)$.

In order to calculate the terms $\int\dg\,\xi_t^*\hat{P}$, we need to prove an extension of Novikov theorem 
for Grassmann Gaussian noise (for the bosonic case, see Ref.~\cite{YDGS99}). In the fermionic case, we have
two kinds of Novikov-type theorems corresponding to the left and right functional derivatives. 
\\
Left type:
\begin{equation}
\int\dg\, \xi^*_t \hat{P} = -\int^t_0 \rd s\, \sum_k \frac{\pa\xi^*_t}{\pa\xi^*_k}\frac{\pa\xi_s}{\pa\xi_k} \int\dg\, \hat{P} \ola{\delta}_{\xi_s};
\end{equation}
\\
Right type:
\begin{equation}
\int\dg\, \hat{P}\xi_t = -\int_0^t \rd s\, \sum_k \frac{\pa\xi_t}{\pa\xi_k}\frac{\pa\xi_s^*}{\pa\xi_k^*} \int\dg\,\ora{\delta}_{\xi_s^*}\hat{P};
\end{equation}
where $\ola{\delta}_{\xi_s^*}$ ($\ora{\delta
}_{\xi_s^*}$) is the right (left) functional derivative with respect to $\xi_s^*$. 
Applying the  Novikov theorem for the Grassmann noise to Eq.~(\ref{rho3}), the formal exact master equation
can be simplified into a compact form
\begin{equation}
\pa_t\hat{\rho}_r = -\ri[\hat{H}_\rs,\hat{\rho}_r] + \{\int\mathcal{D}_g [\bm{\xi}][\bar{Q}\hat{P},\hat{L}\+] + \rm{h.c.}\}, \label{MEQ}
\end{equation}

Similar to the derivation for the SSE, the above derivation for non-Markovian master equation is only valid for the vacuum reservoir, 
in which we assume the system and environment are initially in
the state $|\psi_{\mathrm{tot}}(0)\rangle=|\psi(0)\rangle_{\rm S}\wedge
|\mathrm{vac}\rangle_{\rm R}$. However, the finite temperature case can be easily incorporated
in our approach, as shown in the examples below. 

In a special case where  the fermionic $\hat{Q}$ operator is independent of
noise, the master equation Eq.~(\ref{MEQ}) takes a very simple form,%
\begin{equation}
\partial_{t}\hat{\rho}_{r}=-\ri[\hat{H}_{\mathrm{S}},\hat{\rho}%
_{r}] + \{ [\bar{Q}\hat{\rho_r},\hat{L}^{\dagger}]+\mathrm{h.c.} \}. \label{MEQOfree}%
\end{equation}
If we take the Markovian correlation function $K(t,s)=\Gamma\delta(t,s)$, then
$\bar{Q}$ reduces to $\bar{Q}=\Gamma\hat{L}/2$, and the master equation
will reduce to the standard Lindblad Markov master equation, 
\begin {eqnarray}
 \partial_{t}\hat{\rho}_{r}=-\ri[\hat{H}_{\mathrm{S}},\hat{\rho}_{r}] + \{ \Gamma/2[\hat{L}\hat{\rho}_{r},\hat{L}^{\dagger}]+\mathrm{h.c.} \}.
\end {eqnarray}

%%%%%%%%%%%%%%
\section{Many-fermion system coupled to a vacuum fermionic reservoir} %-
\label{example1}
 The first example considers a many-fermion open system coupled to a fermionic
bath initially in the vacuum state. The total Hamiltonian is 
\beqa
 \hat{H}_{\mathrm{tot}} &=& \sum_{j=1}^{N}\Omega_{j}\hat{d}_{j}^{\dagger}\hat{d}_{j}+\sum_{k}\omega
_{k}\hat{b}_{k}^{\dagger}\hat{b}_{k} \nonumber\\
&&+\sum_{j,k}t_{j,k}\hat{d}_{j}^{\dagger}\hat{b}_{k}+t_{j,k}\hat{b}_{k}^{\dagger}\hat{d}_{j},
\eeqa
  where $\hat{d}_{j}$
and $\hat{d}_{j}^{\dagger}$ ($j=1$ to $N$) are the annihilation and
creation operators of the fermions in the system, and $\hat{b}_{k}$ and $\hat
{b}_{k}^{\dagger}$ are the fermionic annihilation and creation operators for
the bath. Here $\hat{H}_{\mathrm{S}}=\sum_{j}\Omega_{j}\hat{d}_{j}^{\dagger
}\hat{d}_{j}$, and the coupling operator is $\hat{L}=\sum_{j}\hat{d}_{j}$.
Then, the fermionic SSE is given by%
\begin{equation}
\ri\partial_{t}|\psi_{t}\rangle=(\sum_{j}\Omega_{j}\hat{d}_{j}^{\dagger}\hat
{d}_{j}-\ri\sum_{j}\hat{d}_{j} \xi_t^{\ast}-\ri\sum_{j}\hat{d}%
_{j}^{\dagger}\bar{Q})|\psi_{t}\rangle,
\end{equation}
where the $\hat{Q}$ operator is given by $\hat{Q}=\sum_{j}f_{j}(t,s)\hat
{d}_{j}.$ Substituting the $\hat{Q}$ operator into the Eq.~(\ref{EQO}), we
obtain the differential equations for the time-dependent coefficients
$f_{j}(t,s)$ as 
\begin {eqnarray}
 \frac{\partial}{\partial t}f_{j}(t,s)= \ri\Omega_{j}f_{j}(t,s)+\sum_{k=1}^{N}F_{k}(t)f_{j}(t,s)
\end {eqnarray}
 with the final condition
$f_{j}(t,s=t)=1$. $F_{j}(t)$ is defined as $F_{j}(t)=\int_{0}^{t} \rd s\,%
K(t,s)f_{j}(t,s)$. Thus, the exact $\hat{Q}$ operator is fully determined.
Then one immediately obtains the exact master equation,
\begin{equation}
\partial_{t}\hat{\rho}_r =-\ri[\hat{H}_{\mathrm{S}},\hat{\rho}_r]+\{[(\sum_{j=1}%
^{N}F_{j}(t)\hat{d}_{j})\hat{\rho}_r,\sum_{j=1}^{N}\hat{d}_{j}^{\dagger
}]+\mathrm{h.c.}\}.
\end{equation}
As a special case of interest, we consider both the system and the reservoir just containing one mode with equal frequencies (resonant condition). 
Then the exact master equation reduces to 
\begin{equation}
\partial_{t}\hat{\rho}_r =-\ri\Omega [\hat{d}^\dagger \hat{d},\hat{\rho}_r]+\{[g \tan(g t)\hat{d}\hat{\rho}_r,\hat{d}^{\dagger
}]+\mathrm{h.c.}\},
\end{equation}
where $g$ is the coupling constant. This case is an extreme case of the non-Markovian evolution with the non-Markovianity being infinity \cite{Breuer2009}. 
In general, the negative values of the coefficients dictate the features of the
non-Markovian evolution of an open quantum system. A more complete studies on non-Markovian fermonic systems will be conducted in the future
publications \cite{New}.

\section{Single quantum dot (QD) coupled to a finite-temperature fermionic bath} %-----------------------------------------------------------------------------------------------------------
\label{example2}
 To illustrate how to establish a fermionic SSE for the case of finite temperature reservoirs,  for simplicity, we use the example of 
 a single QD coupled to a finite-temperature fermionic bath. As stated before, a realistic generalization to two finite temperature reservoirs is
 straightforward. In the standard Hamiltonian in Eq.~(\ref{Htot}) and Eq.~(\ref{Hint}), we choose $\hat
{H}_{\mathrm{S}}=\omega_{0}\hat{d}^{\dagger}\hat{d}$ and $\hat{L}=\hat{d}$, then the total Hamiltonian is now given by, 
\beq
\hat{H}_\rt =\om_{0}\hat{d}\+\hat{d}+\sum_{k}t_{k}(\hat{d}\+\hat{b}_{k}
 +\hat{b}_{k}\+\hat{d})+\sum_{k}\om_{k}\hat{b}_{k}\+\hat{b}_{k}.
\eeq
It is known that the finite temperature model can be transformed into the
vacuum case by introducing a fictitious reservoir \textquotedblleft$c$" (a ``hole"  system with negative energies), which is decoupled from
the system and reservoir ``$b$'', so the quantum dynamics will not be affected \cite{Strunz,Yu2004}. 
 With the fictitious reservoir  $\hat{H}_{\mathrm{R}}^{fic}=\sum_{k}(-\omega_{k})\hat{c}_{k}^{\dagger}\hat{c}_{k}$,  the total 
 Hamiltonian may be written as,
 \begin {eqnarray}
  \hat{H}_\rt' & =\om_{0}\hat{d}\+\hat{d}+\sum_k t_k(\hat{d}\+\hat{b}_k + \hat{b}_k\+\hat{d}) \non \\
& + \sum_k\om_k(\hat{b}_k\+\hat{b}_k-\hat{c}_k\+\hat{c}_k). \label{ficH}
 \end {eqnarray}
By properly choosing the parameters of bath
\textquotedblleft$c$", the combined bath \textquotedblleft$b+c$" can be
initially prepared in a pure state which is equivalent to the vacuum corresponding to bath \textquotedblleft$b'+c'$".
The relation between the original bath and the transformed baths  are given by the following Bogoliubov  transformations,
{\small 
\begin{align}
\hat{b}_{k}'=\sq{1-\bar{n}_{k}}\,\hat{b}_{k}-\sq{\bar{n}_{k}}\,\hat{c}_{k}\+,  &  \quad \hat{b}_{k}'{}\+=\sq{1-\bar{n}_{k}}\,\hat{b}_{k}\+-\sq{\bar{n}_{k}}\,\hat{c}_{k},\non\\
\hat{c}_{k}'=\sq{1-\bar{n}_{k}}\,\hat{c}_{k}+\sq{\bar{n}_{k}}\,\hat{b}_{k}\+,  &  \quad \hat{c}_{k}'{}\+=\sq{1-\bar{n}_{k}}\,\hat{c}_{k}\++\sq{\bar{n}_{k}}\,\hat{b}_{k};\non\\
\hat{b}_{k}=\sq{1-\bar{n}_{k}}\,\hat{b}_{k}'+\sq{\bar{n}_{k}}\,\hat{c}_{k}'{}\+,  &  \quad \hat{b}_{k}\+=\sq{1-\bar{n}_{k}}\,\hat{b}_{k}'{}\++\sq{\bar{n}_{k}}\,\hat{c}_{k}^{\prime},\non\\
\hat{c}_{k}=\sq{1-\bar{n}_{k}}\,\hat{c}_{k}'-\sq{\bar{n}_{k}}\,\hat{b}_{k}'{}\+,  &  \quad \hat{c}_{k}\+=\sq{1-\bar{n}_{k}}\,\hat{c}_{k}'{}\+-\sq{\bar{n}_{k}}\,\hat{b}_{k}';
\end{align}
}
where $\bar{n}_{k}=\frac{1}{1+\re^{\beta(\hbar\om_{k}-\mu)}}$, and $\mu$ is
the chemical potential.
After tracing out the fictitious bath \textquotedblleft$c$" on the effective vacuum, 
the real bath ``$b$" is then prepared in the initial thermal state, \emph{i.e.}, 
\begin {eqnarray}
 \hat{\rho}_{b}(0)=\mathrm{Tr}_{c}[\hat{\rho}_{bc}(0)]={\exp[{-\frac{\hat{H}_{b}-\mu\hat{N}_b}{k_{B}T}}]}/{Z}
\end {eqnarray}
where $Z=\mathrm{Tr}\exp[{-\frac{\hat{H}_{b}-\mu\hat{N}_b}{k_{B}T}]}$ is the
partition function. In such a way, the finite temperature model can be transformed
into an effective vacuum case whose SSE has already been established in the previous sections.

Now  we define the new coupling strength,
\begin{equation*}
g_{k}\eq\sq{1-\bar{n}_{k}}t_{k},\quad f_{k}\eq\sq{\bar{n}_{k}}t_{k},
\end{equation*}
then the total Hamiltonian takes the following form,
\begin{align}
\hat{H}_\rt'  &  =\om_0\hat{d}\+\hat{d} + \sum_k (g_k\hat{d}\+\hat{b}'_k + f_k\hat{d}\+\hat{c}'_k{}\+ + g_k\hat{b}'_k{}\+\hat{d}  \non\\
&  + f_k\hat{c}'_k\hat{d}) + \sum_k\om_k (\hat{b}'_k{}\+\hat{b}'_k - \hat{c}'_k{}\+\hat{c}_k').
\end{align}
The coherent states for the two baths can be defined as $|\bm{\xi}\ra\eq\prod_{k,l}(1-\xi_{b',k}\hat{b}%
_{k}'{}\+)(1-\xi_{c',l}\hat{c}_{l}^{\prime\dagger
})|\mathrm{vac}\ra_\rr$. Thus,  two independent Grassmann noises are defined, 
\begin {eqnarray}
& & \xi_{b',t}^* \eq -\ri\sum_k g_k \re^{\ri\om_ k t}\xi_{b',k}^*, \non \\
& & \xi_{c',t}^* \eq -\ri\sum_k f_k \re^{-\ri\om_k t}\xi_{c',k}^*.
\end {eqnarray}
Then, the corresponding $\hat{Q}$ ($\bar{Q}$) operators and correlation
functions are defined as%
\begin{eqnarray}
 & & \hat{Q}_{b'}(t,s,\bm{\xi}^*)|\psi_{t}\ra \eq \ora{\delta}_{\xi_{b',s}^*}|\psi_{t}\ra, \non \\
& & \hat{Q}_{c'}(t,s,\bm{\xi}^*)|\psi_{t}\ra \eq \ora{\delta}_{\xi_{c',s}^*}|\psi_{t}\ra, \non \\
& & \bar{Q}_{b'} \eq \int_0^t \rd s\, K_{b'}(t,s)\hat{Q}_{b'}(t,s,\bm{\xi}^*), \non \\
& & \bar{Q}_{c'} \eq \int_0^t \rd s\, K_{c'}(t,s)\hat{Q}_{c'}(t,s,\bm{\xi}^*), \non \\
& & K_{b'}(t,s)  \eq \sum_k g_k^2\, \re^{-\ri\om_k (t-s)}, \non \\
& & K_{c'}(t,s)  \eq \sum_k f_k^2\, \re^{\ri\om_k (t-s)}.
\end{eqnarray}
With the above definitions,  the SSE governing $|\psi_{t}\ra$ can be written as
\begin{equation}
\ri\pa_{t}|\psi_{t}\ra=[\hat{H}_\rs +\ri\hat{d}\+(\xi_{c',t}^* - \bar{Q}_{b'}) +\ri\hat{d}(\bar{Q}_{c'} - \xi_{b',t}^*)]|\psi_t\ra.
\end{equation}
Following the procedure of deriving master equation from the general SSE for the vacuum fermionic bath, 
we obtain the formal exact master equation of the case of finite temperature,
\begin{align}
\pa_t\hat{\rho}_r  &  = -\ri[\hat{H}_\rs,\hat{\rho}_r] + \int\mathcal{D}_g [\bm{\xi}]([\bar{Q}_{b'}\hat{P},\hat{d}\+] \non \\
& +[\hat{d},\bar{Q}_{c'}\hat{P}] + \rm{h.c.}).
\end{align}
For the finite temperature case, the $\hat{Q}$ operators are not noise-free, hence we need to use the Heisenberg approach (see Appendix E) 
to derive the corresponding convolutionless master equation (an example of using Heisenberg approach in
the case of bosonic bath can be found in Ref.~\cite{Strunz-Yu2004,Yu2004}).

The convolutionless master equation takes the following form,
\begin{equation}
\label{MED1d}
\partial_{t}\hat{\rho}_{r}=-\ri[\hat{H}_{\mathrm{S}},\hat{\rho}_{r}%
]\newline+\{F_{1}(t)[\hat{d}\hat{\rho}_{r},\hat{d}^{\dagger}]+F_{2}%
(t)[\hat{\rho}_{r}\hat{d},\hat{d}^{\dagger}]+\text{\textrm{h.c}.}\}
\end{equation}
where the time-dependent coefficients $F_{i}(t)$ ($i=1,2$) are 
\begin{align}
&  F_i(t)=\int_0^t \rd s\,[K_{b'}(t,s)u^{b'}_i(t,s) - K_{c'}^*(t,s)u^{c'}_i(t,s)],
\label{F1F2}%
\end{align}
and $u^\mu_i$ ($i=1,2$, $\mu=b'$ or $c'$) satisfy the
following equations
\begin {widetext}
\begin {align}
\pa_s u^{b'}_i(t,s)  &  = -\ri\om_0u^{b'}_i(t,s)+ [\int_s^t \rd s'\, K_{c'}(s',s)  -\int_0^s \rd s'\, K_{b'}(s,s') ] u^{b'}_i(t,s') + \int_0^t \rd s'\, K_{c'}(s',s)u^{c'}_i(t,s'), \non \\
\pa_s u^{c'}_i(t,s)  &  = -\ri\om_0u^{c'}_i(t,s)+ [\int_s^t \rd s'\, K_{b'}^*(s',s)  -\int_0^s \rd s'\, K_{c'}^*(s,s') ] u^{c'}_i(t,s') + \int_0^t \rd s'\, K_{b'}^*(s',s)u^{b'}_i(t,s'), \label{ubc}%
\end {align}
\end {widetext}
with the final conditions: $u^{b'}_1(t,s=t) = u^{c'}_2(t,s=t) = 1$, and $u^{b'}_2(t,s=t) = u^{c'}_1(t,s=t) = 0$.

\begin{figure}[ptb]
\begin{center}
\includegraphics
[
trim=0.000000in 0.000000in 0.000000in -0.145355in,
height=2.8106in,
width=3.2621in
]
{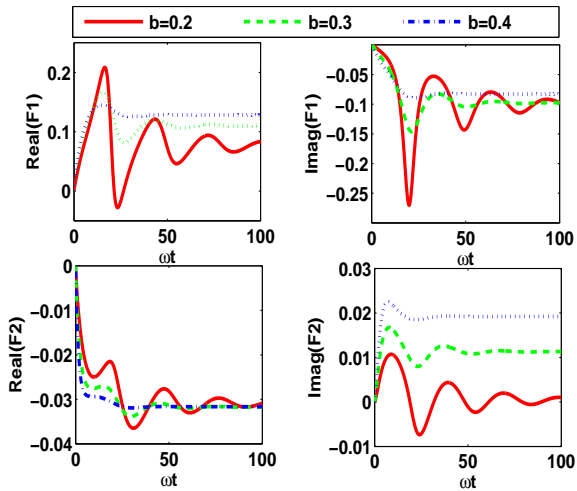}
\end{center}
\caption{Time evolution of the coefficients for the single quantum dot in a finite
temperature bath with different bandwidths. The real (imaginary) part of the
coefficients $F_{1}$ ($F_{2}$) are plotted separately. The parameters are
$T=100{\rm mK}$, $\mu=2\times10^{-5}{\rm eV}$, $\omega_{0}=3\times10^{-5}{\rm eV}$.}%
\label{sd_coeff}%
\end{figure}

\begin{figure}[ptb]
\begin{center}
\includegraphics[
trim=0.000000in 0.000000in -0.055938in 0.000000in,
height=1.6in,
width=3.2621in
]{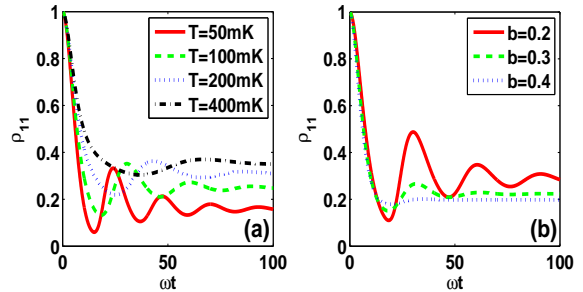}
\end{center}
\caption{Time evolution of $\rho_{11}$ for the single quantum dot with different parameters of the bath.
(a) is plotted with the different temperatures $T$, and (b) is plotted with the different
bandwidths of the spectral density. The other parameters are $\mu
=2\times10^{-5}{\rm eV}$, $\omega_{0}=3\times10^{-5}{\rm eV}$.}%
\label{sd}%
\end{figure}

Generally speaking, the time-dependent coefficients of the exact master equation may evolve
in a complicated way (for example, taking negative values 
as a typical manifestation of non-Markovian behaviors, see \cite{Breuer2009}) and can 
be sensitively affected by the parameters of the environment. To show the temporal behavior of the exact master equation, 
we plot the Fig.~\ref{sd_coeff} for the time-evolution of the coefficients $F_{1}(t)$ and $F_{2}(t)$; for simplicity, we
choose a noise-free $\hat{Q}$ operator in our numerical simulations. The
spectral density is chosen as the Lorentzian form 
$$t_{k}^{2}(\omega_{k})\Delta\omega=\frac{\Gamma b^{2}}{(1-\frac{\omega_{k}}{\omega_{0}})^{2}+b^{2}}.$$
 When the bandwidth $b$ is wide, which corresponds to a white noise situation, 
 the coefficients $F_{1}(t)$ and $F_{2}(t)$ must converge to constants rapidly, approaching the Markov limit
\cite{Wiseman}.  On the contrary, if the bandwidth $b$ is narrow, the
distribution of the spectral density should represent a colored noise case, then we
could expect that the non-Markovian properties (\emph{e.g.}, time-dependent coefficients)
becomes dominant. As a direct result of using different $F_{1}(t)$ and $F_{2}(t)$, we could see that in Fig.~\ref{sd} (b), the density
matrix element $\rho_{11}$ performs differently when it converges to the steady state. The wider the bandwidth $b$ expands the faster the steady state could be approached,
and significant fluctuations would come forth when $b$ is small.
Another parameter that will affect the non-Markovian properties is the
temperature of the bath. As shown in Fig.~\ref{sd}
(a), the non-Markovian behaviors become more dominant in the low temperature
regimes.

\begin{figure}[ptb]
\begin{center}
\includegraphics[
trim=0.000000in 0.000000in 0.000000in -1.148132in,
height=2.5097in,
width=3.2621in
]{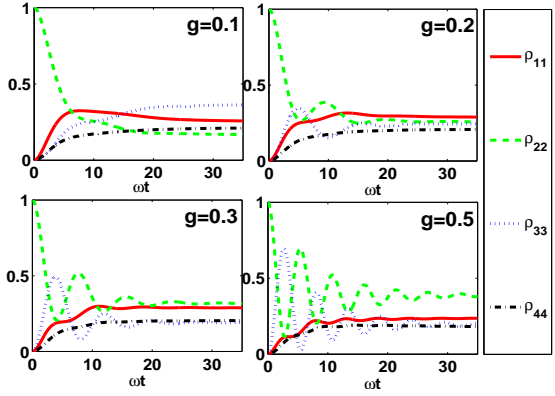}
\end{center}
\caption{Dynamic evolution for the double quantum dot from the initial state $d_{1}^{\dag}|\mathrm{vac}\rangle_{\rm S}$ with
different coupling strength $g$. The other parameters are $T=100{\rm mK}$, $\mu
_{1}=2\times10^{-5}{\rm eV}$, $\mu_{2}=4\times10^{-5}{\rm eV}$, $\omega_{1}=\omega
=2.5\times10^{-5}{\rm eV}$, $\omega_{2}=3.5\times10^{-5}{\rm eV}$.}%
\label{dd}%
\end{figure}

\section{Double QDs coupled to two finite-temperature fermionic baths}%---------------------------------------------------------------------------------------------------------------------
\label{example3}
%%%%%%%%%%%%%%%%%%%%%
 The model considered in this section is more involved, but physically more relevant.  Here, we consider an 
electronic system coupled to two fermionic baths described by the following total Hamiltonian,
\begin{align}
\hat{H}_{\mathrm{tot}}  &  =\omega_{1}\hat{d}_{1}^{\dagger}\hat{d}_{1}+\omega
_{2}\hat{d}_{2}^{\dagger}\hat{d}_{2}+g\hat{d}_{1}^{\dagger}\hat{d}_{2}%
+g^{\ast}\hat{d}_{2}^{\dagger}\hat{d}_{1}\nonumber\\
&  +\sum_{k}\omega_{k}(\hat{b}_{1,k}^{\dagger}\hat{b}_{1,k}+\hat{b}%
_{2,k}^{\dagger}\hat{b}_{2,k})\nonumber\\
&  +\{\sum_{k}t_{2,k}\hat{d}_{2}^{\dagger}\hat{b}_{2,k}+\sum_{k}t_{1,k}\hat
{d}_{1}^{\dagger}\hat{b}_{1,k}+\mathrm{h.c.}\},
\end{align}
where $\hat{d}_{i}$ and $\hat{d}_{i}^{\dagger}$ ($i=1,2$) are the fermionic
annihilation and creation operators of the two quantum dots in the system,
and $\hat{b}_{i,k}$, $\hat{b}_{i,k}^{\dagger}$ are the annihilation and creation operators
for the fermionic baths. This Hamiltonian describes a physical model that double quantum
dots coupled to two fermionic baths with different chemical potentials, 
the \textquotedblleft source" and the \textquotedblleft drain". This model has been 
widely studied by using the fermionic path integral and the input-output approaches
\cite{,ZhangDQD,Search2002}.

Similar to the case of the single QD model discussed before,  the exact SSE for 
the double QDs model can be established as,

\begin{align}
\ri\partial_{t}|\psi_{t}\rangle &  =(\hat{H}_{\mathrm{S}}-\ri\hat{d}_{1}^{\dagger
}\bar{Q}_{1,b^{\prime}}-\ri\hat{d}_{1} \xi_{1,b^{\prime},t}^{\ast}+\ri\hat
{d}_{1}^{\dagger} \xi_{1,c^{\prime},t}^{\ast}+\ri\hat{d}_{1}\bar
{Q}_{1,c^{\prime}}\nonumber\\
&  -\ri\hat{d}_{2}^{\dagger}\bar{Q}_{2,b^{\prime}}-\ri\hat{d}_{2}%
\xi_{2,b^{\prime},t}^{\ast}+\ri\hat{d}_{2}^{\dagger} \xi_{2,c^{\prime
},t}^{\ast}+\ri\hat{d}_{2}\bar{Q}_{2,c^{\prime}})|\psi_{t}\rangle,
\end{align}
where $\bar{Q}_{\mu,\nu}|\psi_t\rangle \equiv \int_{0}^{t} \rd s\, K_{\mu,\nu}%
(t,s)\overrightarrow{\delta}_{\xi_{\mu,\nu,s}^{\ast}}|\psi_t\rangle$ (the
indices  $\mu=1,2$ represent the left and right baths, and $\nu=b^{\prime
}, c^{\prime}$  represent  the fictitious baths $b^{\prime}$ and $c^{\prime}$ in the finite-temperature 
transformation). Therefore, the exact master equation for the double quantum dots system can be derived from the
SSE,
\begin{align}
\pa_t \hat{\rho}_r  &  = -\ri[\hat{H}_\rs , \hat{\rho}_r] + \{ \sum_{j=1}^2 [ (F_1^j(t)\hat{d}_1 + F_2^j(t)\hat{d}_2) \hat{\rho}_r \non \\
   & + \hat{\rho}_r (F_3^j(t)\hat{d}_1 + F_4^j(t) \hat{d}_2)  , \hat{d}_j\+ ] + \rm{h.c.} \}
\end{align}

where
\begin{align}
F^j_i(t)  &  \eq \int_0^t \rd s\, [K_{jb'}(t,s)u^{jb'}_i(t,s) - K_{jc'}^*(t,s)u^{jc'}_i(t,s)],\non\\
\end{align}
where $K_{j\mu}$ ($j=1,2$, $\mu=b'$ or $c'$) are
the correlation functions, and the equations for the coefficients
$u^\mu_i$ ($i=1,2,3,4$, $\mu=1b',2b', 1c' $ or $2c'$) are 
%%%%%%%%%%%%%%%%%%%%%%%%%%%
\begin {widetext}{\small
\begin {align}
  \pa_s u^{1b'}_j(t,s) & = -\ri\om_1 u^{1b'}_j(t,s) -\ri g\, u^{2b'}_j(t,s) + [ \int_s^t \rd s'\, K_{1c'}(s',s) - \int_0^s \rd s'\, K_{1b'}(s,s') ] u^{1b'}_j(t,s') 
       + \int_0^t \rd s'\, K_{1c'}(s',s) u^{1c'}_j(t,s') \non\\
  \pa_s u^{2b'}_j(t,s) & = -\ri\om_2 u^{2b'}_j(t,s) -\ri g^* u^{1b'}_j(t,s) + [ \int_s^t \rd s'\, K_{2c'}(s',s) - \int_0^s \rd s'\, K_{2b'}(s,s') ] u^{2b'}_j(t,s') 
       + \int_0^t \rd s'\, K_{2c'}(s',s) u^{2c'}_j(t,s') \non\\
  \pa_s u^{1c'}_j(t,s) & = -\ri\om_1 u^{1c'}_j(t,s) -\ri g\, u^{2c'}_j(t,s) + [ \int_s^t \rd s'\, K_{1b'}^*(s',s) - \int_0^s \rd s'\, K_{1c'}^*(s,s') ] u^{1c'}_j(t,s')  
       + \int_0^t \rd s'\, K_{1b'}^*(s',s) u^{1b'}_j(t,s') \non\\
  \pa_s u^{2c'}_j(t,s) & = -\ri\om_2 u^{2c'}_j(t,s) -\ri g^* u^{1c'}_j(t,s) + [ \int_s^t \rd s'\, K_{2b'}^*(s',s) - \int_0^s \rd s'\, K_{2c'}^*(s,s') ] u^{2c'}_j(t,s') 
       + \int_0^t \rd s'\, K_{2b'}^*(s',s) u^{2b'}_j(t,s'), \non\\
\end {align}}
\end {widetext}
with the final conditions: $u^{1b'}_1(t,s=t)=u^{2b'}_2(t,s=t)=u^{1c'}_3(t,s=t)=u^{2c'}_4(t,s=t)=1$, and the others are zero.
We omit the mathematical details of solving this model, 
since the procedure is complicated however the main idea is still same to the single QD case.

Here, we only show some properties of this model by plotting the time evolution of the population; and the
detailed study of this model will be discussed elsewhere \cite{New}. In
Fig.~\ref{dd}, we plot the dynamic evolution of the probabilities of all the four
states with different coupling strength between the two QDs. In a long-time limit, 
the system trends to converge to a steady state.
When $t$ is small, the electron tunneling from one dot to the other can be
significantly enhanced by the direct couplings
between the two QDs.

%%%%%%%%%%%%%%%%%%%%%%%%%%%

\section{conclusion}
\label{conclusion}
In this paper, we have developed an exact fermionic stochastic
Schr\"{o}dinger equation approach for solving the quantum open system coupled
to a fermionic environment. The fundamental dynamic equation is derived
directly from the microscopic quantum model without any approximations. By
using the Grassmann noise, the stochastic Schr\"{o}dinger equation approach is
expanded from bosonic to fermionic environments.  Three examples are presented
to show the power of this approach. It is worth noting that the stochastic
Schr\"{o}dinger equation is versatile enough to deal with a generic fermionic
environment incorporating cases from strong system-reservoir interaction to structured reservoirs.
 The exact stochastic approach can be applied to more realistic models when the
approximation methods are used \cite{New}.

{\it Note Added:}  After completion of this work, we became aware of an independent work 
by M. Chen and J. Q. You \cite{Chen-You2012}, who also derived a stochastic diffusive equation by 
using Grassmann coherent state approach. 

\section*{acknowledgements}
We acknowledge the grant support  from the NSF PHY-0925174 and
the AFOSR No. FA9550-12-1-0001.

\appendix

\section{Fermionic coherent state and Grassmann noise}
Fermionic coherent state:  For any set $\bm{\xi}=\{\xi_j\}$ of independent Grassmann numbers, we define the fermionic coherent state $|\bm{\xi}\ra$ as 
\begin {eqnarray}
 |\bm{\xi}\ra \eq \prod_k(1-\xi_k \hat{b}_k\+)|\mathrm {vac}\ra_\rr.
\end {eqnarray}
By our definition, it is easy to verify the results below,
\begin {eqnarray}
 \hat{b}_j|\bm{\xi}\ra=\xi_j|\bm{\xi}\ra=|\bm{\xi}\ra\xi_j, \quad 
      \hat{b}\+_j|\bm{\xi}\ra=-\ora{\pa}_{\xi_j}|\bm{\xi}\ra=|\bm{\xi}\ra\ola{\pa}_{\xi_j}. \non \\
\end {eqnarray}
We can also check the validity of the completeness of the fermionic coherent states,
\begin {eqnarray}
 \hat{I}=\int\dg |\bm{\xi}\,\ra\la\bm{\xi}\,| = \prod_{k}\int \rd \xi_{k}^* \, \rd \xi_{k} \, \re^{-\xi_{k}^*\xi_k}|\bm{\xi}\,\ra\la\bm{\xi}\,|, \non \\ \label{Id}% 
\end {eqnarray}
 where $\dg$ is the Grassmann Gaussian measure.

From the main text of this paper we know that the Grassmann Gaussian noises are generated as:
\begin {eqnarray}
 \xi_t \eq \ri\sum_k t_k \re^{-\ri\om_k t}\xi_k, \quad \xi_s^* \eq -\ri\sum_k t_k \re^{\ri\om_k s}\xi_k^*,
\end {eqnarray}
through these definitions of the noises, we can calculate the correlation function as:
\begin {eqnarray}
 \mm [\xi_t \xi_s^*] = \sum_k t_k^2 \re^{-\ri\om_k (t-s)} \mm [\xi_k \xi_k^*].  
\end {eqnarray}
It is easy to check that  $\mm [\xi_k \xi_k^*]=\int\dg\xi_k \xi_k^*=1$, so 
 $ \mm [\xi_t \xi_s^*]=K(t,s)=\sum_{k}|t_{k}%
|^{2}\re^{-\ri\om_{k}(t-s)}$.

Since time is a continuous index, so we can introduce the functional derivatives with respect to the Grassmann noise and the Taylor expansion of the Grassmann noise functional.
Another useful feature is the chain rule; while deriving the SSE we only need to use a special case of the chain rule 
$\ora{\pa}_{\xi_k} = \int_0^t \rd s \, \frac{ \pa_{\xi_s}  } { \pa_{\xi_k} } \ora{\delta}_{\xi_s} $ and
 $\ola{\pa}_{\xi_k} = \int_0^t \rd s \,\frac{ \pa_{\xi_s}  } { \pa_{\xi_k} } \ola{\delta}_{\xi_s} $. The validity of this special chain rule is easy to check.

\section{Derivation of the equation for fermionic $\hat{Q}$ operator}

%%%%%%%%%%%%%%%%%%%%%%%%%%%%%%%%
Applying the SSE Eq.~(\ref{SSE}) to both sides of the consistency condition,%
\begin{equation}
\ora{\delta}_{\xi_{s}^*}\pa_{t}|\psi_{t}\ra
=\pa_{t}\ora{\delta}_{\xi_{s}^*}|\psi_{t}\ra,
\end{equation}
we will obtain
\begin{align}
\pa_t(\hat{Q})|\psi_{t}\ra &  =(\ora{\delta}_{\xi_s^*}\pa_t-\hat{Q}\pa_t)|\psi_t\ra \non \\
&  =(-\ri\ora{\delta}_{\xi_s^*}\hat{H}_{\rm eff} +\ri\hat{Q}\hat{H}_{\rm eff})|\psi_t\ra \non \\
&  =\{-\ri[\hat{H}_{\rm eff},\hat{Q}] -\ri \ora{\delta}_{\xi_s^*}(\hat{H}_{\rm eff}-\hat{H}_\rs) \} |\psi_t\ra. \label{Qdynamics1}%
\end{align}
So, the dynamics of $\hat{Q}$ operator is %
\begin{equation}
\pa_t \hat{Q} = -\ri[\hat{H}_{\rm eff},\hat{Q}] - \ri \ora{\delta}_{\xi_s^*}(\hat{H}_{\rm eff}-\hat{H}_\rs). 
\end{equation}

\section{Recovering a reduced density operator from the Grassmann trajectories}

First, note that  the following two formulas for the Grassmann variables are necessary for our derivation.

1. For any two Grassmann functions $X(\bm{\xi}\,)$ and $Y(\bm{\xi}\,)$,
one has
\begin{align}
X(\bm{\xi}\,)Y(\bm{\xi}\,)  &  =\frac{1}{2}[Y(\bm{\xi}\,)X(\bm{\xi}\,)+Y(-\bm{\xi}\,)X(\bm{\xi}\,)\non\\
&  +Y(\bm{\xi}\,)X(-\bm{\xi}\,)-Y(-\bm{\xi}\,)X(-\bm{\xi}\,)]. \label{XY}%
\end{align}

2. The resolution of identity for the fermionic coherent states is
given by, 
\begin{equation}
\hat{I}_\rt=\sum_\bn\int\dg\, |\bn\ra_\rs|\bm{\xi}\,\ra\la\bm{\xi}\,|\la\bn|_\rs.  \label{ID}%
\end{equation}

We start the derivation from a lemma \cite{New} without its proof 
\begin {eqnarray}
 \hat{\rho}_r \eq \sum_{\bn ,\, \bl, \, \bmm}|\bn\ra_\rs  \la \bl|_\rr \la \bn|_\rs \hat{\rho}_\rt |\bmm\ra_\rs |\bl\ra_\rr  \la\bmm|_\rs. \label{rho}
\end {eqnarray}
Inserting Eq.~(\ref{ID}) into Eq.~(\ref{rho}), and using Eq. ~(\ref{XY}) to exchange some terms, we can get the reduced density operator in the fermionic coherent state representation.
\begin {widetext}
\begin {eqnarray*}
 \hat{\rho}_r &=& \int \dg \sum_{\bn ,\, \bl, \, \bmm,\, \bn'} |\bn\ra_\rs  \la \bl|_\rr \la \bn|_\rs  |\bn'\ra_\rs|\bx\ra \la\bx| \la\bn'|_\rs \hat{\rho}_\rt |\bmm\ra_\rs |\bl\ra_\rr  \la\bmm|_\rs \\
  &=& \int \dg \sum_{\bn ,\, \bl, \, \bmm,\, \bn'} \frac{1}{2} (|\bn\ra_\rs  \la\bx| \la\bn'|_\rs \hat{\rho}_\rt |\bmm\ra_\rs |\bl\ra_\rr \la \bl|_\rr \la \bn|_\rs  |\bn'\ra_\rs|\bx\ra \la\bmm|_\rs \\
  & & + |\bn\ra_\rs  \la -\bx| \la\bn'|_\rs \hat{\rho}_\rt |\bmm\ra_\rs |\bl\ra_\rr \la \bl|_\rr \la \bn|_\rs  |\bn'\ra_\rs|\bx\ra \la\bmm|_\rs  
      + |\bn\ra_\rs  \la\bx| \la\bn'|_\rs \hat{\rho}_\rt |\bmm\ra_\rs |\bl\ra_\rr \la \bl|_\rr \la \bn|_\rs  |\bn'\ra_\rs|-\bx\ra \la\bmm|_\rs \\
  & & - |\bn\ra_\rs  \la -\bx| \la\bn'|_\rs \hat{\rho}_\rt |\bmm\ra_\rs |\bl\ra_\rr \la \bl|_\rr \la \bn|_\rs  |\bn'\ra_\rs|-\bx\ra \la\bmm|_\rs ).
\end {eqnarray*}
 Using some trick of changing the integration variables, these four terms can be merged into only one term.
 
 \begin {eqnarray*}
  \hat{\rho}_r   &=& \int \dg \sum_{\bn ,\, \bl, \, \bmm,\, \bn'}   
       |\bn\ra_\rs  \la\bx| \la\bn'|_\rs \hat{\rho}_\rt |\bmm\ra_\rs |\bl\ra_\rr \la \bl|_\rr \la \bn|_\rs  |\bn'\ra_\rs|-\bx\ra \la\bmm|_\rs \\
  &=&      \int \dg \sum_{\bn ,\, \bl, \, \bmm}   
       |\bn\ra_\rs  \la\bx| \la\bn|_\rs \hat{\rho}_\rt |\bmm\ra_\rs |\bl\ra_\rr \la \bl|_\rr |-\bx\ra \la\bmm|_\rs.
 \end {eqnarray*}
 \end {widetext}
 In the last row of the above formula, the term $ \sum_{ \bl} |\bmm\ra_\rs |\bl\ra_\rr \la \bl|_\rr |-\bx\ra$ actully equals $|\bmm\ra_\rs |-\bx\ra$. (it is true, but we do not want to provide the details)
 That means the density operator could be written as 
 \begin {eqnarray*}
  \hat{\rho}_r   &=& \int \dg \sum_{\bn ,\, \bmm}    |\bn\ra_\rs  \la\bx| \la\bn|_\rs \hat{\rho}_\rt |\bmm\ra_\rs |-\bx\ra \la\bmm|_\rs
 \end {eqnarray*}
 After using a short hand notation, the density operator shows us the familiar form \cite{Strunz,YDGS99}
 \begin {eqnarray}
  \hat{\rho}_r   = \int \dg\,  \la\bx| \hat{\rho}_\rt  |-\bx\ra = \int \dg\, |\psi_t\ra\la\psi_t^-|, \label{rho2}
 \end {eqnarray}
 where $\hat{\rho}_\rt \eq |\psi_\rt^I(t)\ra\la\psi_{\rt}^I(t)|$ is the density operator for a pure state of the total system, and $\la\psi_t^-|\eq \la \psi_t(-\bx)|$ is a 
 quantum trajectory corresponding to a noise $-\xi_t$.

\section{Proof of the extended Novikov theorem}
 We first separate the Grassmann Gaussian measure into two parts: Grassmann measure part and Gaussian part, then give them new notations
 $$\mathcal D [\bm\xi] \equiv \prod_k \rd\xi^*_k \cdot \rd\xi_k, \quad G(\bx) \equiv \prod_k \re^{-\xi^*_k \cdot \xi_k}. $$
 Then we can prove the left-Novikov theorem:
 \begin {eqnarray}
   \int \dg\, \xi^*_t \hat{P} &=& \sum_k \fr{\pa \xi^*_t}{\pa \xi^*_k}\int\, \mathcal D [\bm\xi] G(\bx)\, \xi^*_k \hat{P} \non \\
    &=& \sum_k \fr{\pa \xi^*_t}{\pa \xi^*_k}\int\,\mathcal D [\bm\xi] \ora{\pa}_{\xi_k} G(\bx)\,  \hat{P} \non \\
    &=& \sum_k \fr{\pa \xi^*_t}{\pa \xi^*_k}\int\,\mathcal D [\bm\xi] \{ \ora{\pa}_{\xi_k} [G(\bx)  \hat{P}] \non \\
    & & -\ora{\pa}_{\xi_k} \hat{P}\,  G(\bx)  \} \non \\
    &=& -\sum_k \fr{\pa \xi^*_t}{\pa \xi^*_k}\int\,\mathcal D [\bm\xi] \ora{\pa}_{\xi_k} \hat{P}\,  G(\bx)   \non \\
    &=& \sum_k \fr{\pa \xi^*_t}{\pa \xi^*_k}\idg\, \hat{P}\,\tilde{}\ola{\pa}_{\xi_k} \non \\
    &=& \int^t_0 \rd s  \sum_k\fr{\pa \xi^*_t}{\pa \xi^*_k}\fr{\pa \xi_s}{\pa \xi_k}
       \idg\, \hat{P}\,\tilde{}\ola{\delta}_{\xi_s} \non \\
 \end {eqnarray}
 From the third row to fourth row we use some special features of Grassmann variables
 $$\int \rd\xi_k = \ora{\pa}_{\xi_k}, \quad \ora{\pa}_{\xi_k}\ora{\pa}_{\xi_k} = 0.$$
 In the fifth row `` $\tilde{}$ '' is one kind of fermionic parity operation, and under the even state assumption, we have the conclusion $\hat{P}\,\tilde{} = \hat{P}$. 
 Right-Novikov theorem can be proved similarly.

\section{Heisenberg approach and convolutionless master equation}

In the interaction picture, the dynamics of an operator $\hat{a}$ is
\begin{equation}
\pa_t \hat{a}(t) = \ri \re^{\ri\hat{H}_\rt t} \re^{-\ri\hat{H}_{\rm R}t} [\hat{H}_\rt^{I}(t),\hat{a}] \re^{\ri\hat{H}_{\rm R}t} \re^{-\ri\hat{H}_\rt t}.
\end{equation}
For the single QD model,
\begin{align}
\hat{H}_\rt^{I}  &  = \om_0\hat{d}\+\hat{d} + \sum_k (g_k \re^{-\ri\om_k t}\hat{d}\+\hat{b}'_k + f_k \re^{-\ri\om_k t}\hat{d}\+\hat{c}'_k{}\+ \non \\
&  + g_k \re^{\ri\om_k t}\hat{b}'_k{}\+\hat{d} + f_k \re^{\ri\om_k t}\hat{c}'_k\hat{d}).
\end{align}
Thus, the evolution of the reservoir operators are
\begin{align}
\hat{b}'_k (s)  &  = \hat{b}'_k + \int_0^s \rd s'\, \frac{\pa\xi_{b',s'}^*}{\pa\xi_{b',k}^*}\hat{d}(s'), \non \\
\hat{c}'_k (s)  &  = \hat{c}'_k - \int_0^s \rd s'\, \frac{\pa\xi_{c',s'}^*}{\pa\xi_{c',k}^*}\hat{d}\+(s'), \non 
\end {align}
\begin {align}
\hat{b}'_k{}\+(s)  &  = \hat{b}'_k{}\+(t) - \int_s^t \rd s'\, \frac{\pa\xi_{b',s'}}{\pa\xi_{b',k}}\hat{d}\+(s'), \non \\
\hat{c}'_k{}\+(s)  &  = \hat{c}'_k{}\+(t) + \int_s^t \rd s'\, \frac{\pa\xi_{c',s'}}{\pa\xi_{c',k}}\hat{d}(s').
\end{align}
Define $\hat{U}(t) \eq \re^{\ri\hat{H}_{\rm R}t} \re^{-\ri\hat{H}_\rt t}$, then we can prove
\begin{align}
\la\bm{\xi}|\hat{U}(t)\hat{d}(s)|\psi_\rt(0)\ra &  =\hat{Q}_{b'}(t,s,\bm{\xi}^*)|\psi_{t}\ra, \non \\
\la\bm{\xi}|\hat{U}(t)\hat{d}\+(s)|\psi_\rt(0)\ra &  =-\hat{Q}_{c'}(t,s,\bm{\xi}^*)|\psi_{t}\ra. \label{dQcorrespondence}%
\end{align}
Using these relations, we could derive a set of closed differential equations
with respect to the time $s$.

Define%
\begin{align}
\hat{R}_{b'}(t,s)  &  \eq \int\mathcal{D}_g [\bm{\xi}]\, \hat{Q}_{b'}(t,s,\bm{\xi}_{b'}^*)\hat{P}, \non \\
\hat{R}_{c'}(t,s)  &  \eq \int\mathcal{D}_g [\bm{\xi}]\, \hat{Q}_{c'}(t,s,\bm{\xi}_{c'}^*)\hat{P},
\end{align}
then, the equations for $\hat{R}_{b'}$ and $\hat{R}_{c'}$  are given by%
\begin {widetext}
\begin{align}
\pa_s\hat{R}_{b'}(t,s)  &  = -\ri\om_0\hat{R}_{b'}(t,s)-\int_0^s \rd s'\, K_{b'}(s,s')\hat{R}_{b'}(t,s') + \int_s^t \rd s'\, K_{c'}(s',s)\hat{R}_{b'}(t,s')
  - \int_0^t \rd s'\, K_{c'}^*(s,s')\hat{R}\+_{c'}(t,s'). \non \\
\pa_s\hat{R}_{c'}(t,s)  &  = \ri\om_0\hat{R}_{c'}(t,s)-\int_0^s \rd s'\, K_{c'}(s,s')\hat{R}_{c'}(t,s') + \int_s^t \rd s'\, K_{b'}(s',s)\hat{R}_{c'}(t,s')
  - \int_0^t \rd s'\, K_{b'}^*(s,s')\hat{R}\+_{b'}(t,s'). \label{psRbc}%
\end{align}
\end {widetext}
The solutions of $\hat{R}_{b'}$ and $\hat{R}_{c'}$ may be written as
\begin{align}
\hat{R}_{b'}(t,s)  &  = u^{b'}_1(t,s) \hat{d} \hat{\rho}_{r}(t) + u^{b'}_2(t,s) \hat{\rho}_{r}(t) \hat{d},\non\\
-\hat{R}_{c'}\+ (t,s)  &  = u^{c'}_1(t,s) \hat{d} \hat{\rho}_{r}(t) + u^{c'}_2(t,s) \hat{\rho}_{r}(t) \hat{d}. \label{ExpandR}%
\end{align}
Substitution of  Eq.~(\ref{ExpandR}) into Eq.~(\ref{psRbc}), the
equations for $u^\mu_i$ ($i=1,2$, $\mu=b'$ or $c'$) will be
derived just as it is shown in Eq.~(\ref{ubc}). Then the exact master equation can be
derived given by Eq.~(\ref{MED1d}).


\begin{thebibliography}{99}       

\bibitem {xx} U. Weiss, \emph{Quantum Dissipative Systems} (World Scientific,
Singapore, 1999).


\bibitem {yy} M. Di Ventra, \emph{Electrical Transport in Nanoscale Systems}
(Cambridge University Press, Cambridge, 2008).

\bibitem {nnn} M. Di Ventra, and R. D'Agosta, Phy.  Rev. Lett.  {\bf 98}, 226403 (2007); 
R. D'Agosta, and M. Di Ventra, Phys. Rev. B {\bf 78}, 165105 (2008).

\bibitem {mmm} E. A. Calzetta, and B. L. Hu, \emph{Nonequilibrium Quantum Field
Theory} (Cambridge University Press, New York, 2008).

\bibitem {cc} J. Q. You, and F. Nori,  Nature {\bf 474}, 589 (2011). 

\bibitem {Hu0} H. P. Breuer, and F. Petruccione, {\em The Theory of Open Quantum Systems}
(Oxford University Press, USA, 2002).

\bibitem {Breuer2009}  For more rigorous discussions on non-Markovianity, see,  H. P. Breuer, E. M. Laine, and J. Piilo, Phys. Rev. Lett.
\textbf{103}, 210401 (2009); \'{A}. Rivas, S. F. Huelga, and M. B. Plenio, Phys. Rev. Lett.  \textbf{105}, 050403 (2010).

\bibitem {Feynman-Vernon}R. P. Feynman, and F. L. Vernon, Ann. Phys.
\textbf{24}, 118 (1963).

\bibitem {Leggett}A. O. Caldeira, and A. J. Leggett, Physica A \textbf{121},
587 (1983).

\bibitem {Hu1}B. L. Hu, J. P. Paz, and Y. Zhang,  Phys. Rev. D \textbf{45},
2843 (1992).

\bibitem {Diosi1}L. Di\'osi, and W. T. Strunz, {\ Phys.\ Lett. A\/}
\textbf{235}, 569 (1997).

\bibitem {Diosi2}L. Di\'osi, N. Gisin, and W. T. Strunz, {\ Phys.\ Rev. A\/}
\textbf{58}, 1699 (1998).

\bibitem {Strunz}W. T. Strunz, L. Di\'osi, and N. Gisin, {\ Phys.\ Rev.\ Lett.}
\textbf{82}, 1801 (1999).

\bibitem {YDGS99}T. Yu, L. Di\'osi, N. Gisin,  and W. T. Strunz,
{\ Phys.\ Rev. A\/} \textbf{60}, 91 (1999).

\bibitem {Jing}J. Jing,  and T. Yu, {\ Phys.\ Rev.\ Lett.} \textbf{105}, 240403 (2010).

\bibitem {Strunz-Yu2004} W. T. Strunz,  and T. Yu, {\ Phys.\ Rev. A\/}
\textbf{69}, 052115 (2004).

\bibitem {Yu2004} T. Yu, Phys. Rev. A \textbf{69}, 062107 (2004).

\bibitem {Gisin} N. Gisin,  and I. C. Percival, J. Phys. A \textbf{25}, 5677 (1992);
\textbf{26}, 2233 (1993).

\bibitem {Plenio} M. B. Plenio,  and  P. L. Knight,  Rev. Mod. Phys. {\bf 70}, 101 (1998).

\bibitem {Quantum Noise} C. W. Gardiner, and P. Zoller, \emph{Quantum Noise}
(Springer-Verlag, Berlin, 2004).


\bibitem {Meir} Y. Meir, N. S. Wingreen, and P. A. Lee, Phys. Rev. Lett. {\bf 66}, 3048 (1991).

\bibitem {Meir2} Y. Meir, and N. S. Wingreen, Phys. Rev. Lett. {\bf 68}, 2512 (1992). 

% \bibitem {Jauho} A. P. Jauho {\it et al}, Phys. Rev. B {\bf 50}, 5528 (1994).
\bibitem {Jauho}
A. P. Jauho, N. S. Wingreen, and Y. Meir, Phys. Rev. B {\bf 50}, 5528 (1994).
\bibitem {Wingreen} N. S. Wingreen and Y.  Meir Phys. Rev. B {\bf 49}, 11040 (1994). 

\bibitem {Fransson} J. Fransson,  Phys. Rev. B  {\bf 72}, 075314 (2005). 


\bibitem {Zwanzig} R. Zwanzig,  J.  Chem. Phys. {\bf 33}, 1338 (1960).

\bibitem {Ting} L. Y. Chen, and C. S. Ting, Phys. Rev. B \textbf{43}, 4534 (1991).

\bibitem {ZhangDQD} M. W. Y. Tu,  and W.-M. Zhang, Phys. Rev. B \textbf{78},
235311 (2008); M. W.-Y. Tu, M.-T. Lee, and W.-M. Zhang, Quant. Inf. Process \textbf{8}, 631 (2009); J. Jin, M. W.-Y. Tu, 
W.-M. Zhang,  and Y. Yan, New J. Phys. {\bf 12}, 083013 (2010).

\bibitem {Wiseman} H.-S. Goan, G. J. Milburn, H. M. Wiseman, and H. B. Sun,
Phys. Rev. B \textbf{63}, 125326 (2001); H.-S. Goan, and G. J. Milburn, Phys.
Rev. B \textbf{64}, 235307 (2001).


\bibitem {XinyuFB}  An interesting case where the system commutes with the fermionic
(effective) bath has been considered in X. Zhao, W. Shi, L.-A. Wu, and T. Yu, 
Phys. Rev. A  {\bf 86}, 032116 (2012).


\bibitem {Berezin}F. A. Berezin, \emph{The Method of Second Quantization}
(Academic Press, New York and London, 1966).

\bibitem {Glauber1999}K. E. Cahill, and R. J. Glauber, Phys. Rev. A
\textbf{59}, 1538 (1999).

\bibitem {Strunz1996} W. T. Strunz, Phys. Lett. A {\bf 224}, 25 (1996). 

\bibitem {Wufu} W. Shi and T. Yu, unpublished (2013).

\bibitem {Corn}  B. Corn, J. Jing, and T. Yu,  Submitted to Phys. Rev. A.

\bibitem {New} W. Shi, X. Zhao,  and T. Yu,  unpublished.


\bibitem {Search2002} C. P. Search, S. P\"{o}tting, W. Zhang, and P. Meystre,
Phys. Rev. A \textbf{66}, 043616 (2002).


\bibitem {Chen-You2012}  M. Chen, and J. Q. You,  arXiv:1203.2217.

\end{thebibliography}
\end{document}